\documentclass[twocolumn,amssymb,nobibnotes,aps,pra,showpacs]{revtex4}
  \usepackage[dvips]{color}
  \usepackage{newlfont}
  \usepackage{graphicx}
  \usepackage{amssymb}
  \usepackage{amsmath}
  \usepackage[latin1]{inputenc}
  \usepackage{float}

\begin{document}

\title{Correspondence principle for the diffusive dynamics of a quartic oscillator: deterministic aspects and the role of temperature}
\author{Renato M. Angelo}
\email{renato@fisica.ufpr.br}
\affiliation{Federal University of Paraná, Department of Physics, P. O. Box 19044, Curitiba, 81531 990, Paraná, Brazil.}

\begin{abstract}
The correspondence principle is investigated in the framework of deterministic predictions for individual systems. Exact analytical results are obtained for the quantum and Liouvillian dynamics of a nonlinear oscillator coupled to a phase-damping reservoir at a finite temperature. In this context, the time of critical wave function spreading -- the Ehrenfest time -- emerges as the characteristic time scale within which the concept of deterministic behavior is admissible in physics. A scenario of {\em quasi-determinism} may be then defined within which the motion is experimentally indistinguishable from the truly deterministic motion of Newtonian mechanics. Beyond this time scale, predictions for individual systems can be given only statistically  and, in this case, it is shown that diffusive decoherence is indeed a necessary ingredient to establish the quantum-classical correspondence. Moreover, the high-temperature regime is shown to be an additional condition for the quantum-classical transition and, accordingly, a lower bound for the reservoir temperature is derived for our model.


\end{abstract}

\maketitle

\section{Introduction}

A widely accepted explanation for the emergence of classical behavior from the quantum substratum is provided by the environment-induced decoherence (EID) program \cite{zurek03,omnes,joos,isar}: the interaction of the system with many uncontrollable external degrees of freedom (the environment) yields a dramatic destruction of quantum coherences prohibiting, as a consequence, the existence of quantum superpositions in the macroscopic world. Decoherence, an experimentally demonstrated effect \cite{brune,cheng,myatt} to which real systems are inevitably submitted, is the essence of the arguments given by EID defenders to explain, for instance, why an astronomical body behaves classically \cite{zurek98,zurek96} and to address fundamental questions, as for instance the Schröedinger cat paradox and the measurement problem \cite{omnesB}.

Recent works, however, have discussed the real effectiveness of decoherence as a necessary mechanism to guarantee the quantum-classical transition. In reference \cite{wiebe}, e.g., the authors show that a small amount of smoothing (due to apparatus resolution) is sufficient to ensure the classical limit. In \cite{adelcio}, it is argued that diffusive decoherence can produce only an attenuation of quantum coherences, not their complete destruction. The authors thus conclude that the classical limit is a matter of experimental resolution. Another important objection raised in references \cite{wiebe,ballentine} points out to the fact that decoherence is not able to reduce the spreading of the wave function. Yet, in reference \cite{petitjean} relevant questions concerning the role of the high-temperature limit and the generality of environment models have been addressed. Such a present debate forbids one to believe that the understanding of the quantum-classical transition is an exhausted subject.

In this work the correspondence principle is discussed by means of an analytical analysis carried out for the the dynamics of a nonlinear oscillator coupled with a phase-damping reservoir composed by $N$ harmonic oscillators in thermal equilibrium. Such a thermal bath has been shown to be effective in modelling nondissipative decoherence for a long range of finite temperatures \cite{angelo06}. With the help of this model, the quantum-classical correspondence is analyzed in two main directions, as indicated below.

(i) On one hand, it is investigated the correspondence between the
quantum centroid and the Newtonian trajectory. This analysis is
motivated by Einstein's paper to Born about the foundations of
quantum mechanics \cite{einstein}. Einstein concludes that the only
possible interpretation for the wave function is the one based on
statistical ensembles but, on the other hand, emphasizes the
``inevitably conception that Physics must ferment a realistic
description of only one system''. Indeed, ``nature as a whole may be
thought as an individual system (existing only one time, with no
need for repetitions) and not as a `ensemble of
systems'.''\cite{einstein}. Einstein's dissatisfaction is invoked here in
terms of the following motivating question: {\em what can quantum
mechanics predict about the future of an individual system?}
Obviously, such a question intends to challenge the quantum
formalism to realize a task which is supposed to be an exclusiveness
of Newtonian mechanics. However, since physics is fundamentally concerned with experimental observations, which in turn are inevitably limited by the apparatus resolution, it is reasonable to conceive that neither the idea of an objective reality (usually associated with classical beliefs) nor a completely indeterministic scenario (usually associated with quantum world) can be maintained. In fact, as it is argued in Sec.\ref{QNC}, once a margin of error is defined the concept of deterministic behavior can be physically accepted, at least for the short-time dynamics of initially localized wave functions. Moreover, it is shown that within such a short-time regime (the Ehrenfest regime) decoherence is not effective, in agreement with the claims of ref.\cite{wiebe,ballentine}.

(ii) On the other hand, beyond this time scale within which deterministic behavior is expected to be experimentally verifiable, predictions for individual systems can be given only in terms of statistical averages and variances. In this case, the quantum-classical correspondence is investigated in terms of comparisons between quantum and Liouvillian results, for which the experimental resolution is also regarded. In Sec.\ref{QLC}, it is shown that decoherence is indeed a necessary mechanism to promote the quantum-classical correspondence. Our results show in addition that the reestablishment of the quantum-classical correspondence can be ensured by purely diffusive decoherence only within the regime of high temperatures.

\section{The model}
\label{model}

Recently, a model of reservoir has been proposed which is able to describe nondissipative decoherence at finite temperatures \cite{angelo06}. Here, it is investigate the dynamics of a system composed by such a reservoir coupled to the well known \cite{milburn,lili,adelcio} quartic oscillator. The Hamiltonian of the model reads
\begin{subequations}
\begin{eqnarray}
\hat{H}&=&\hat{H}_S+\hat{H}_R+\hat{H}_I,\label{Htot}
\end{eqnarray}
where
\begin{eqnarray}
\hat{H}_S&=&\hbar\,\omega_s\,\left(\hat{n}_s+\frac{1}{2}\right)+\hbar^2\,g_s\,\left(\hat{n}_s+\frac{1}{2}\right)^2, \label{HHO}\\
\hat{H}_R&=&\sum\limits_{k=1}^N\hbar\,\omega_k\,\left(\hat{n}_k+\frac{1}{2}\right), \label{HE}\\
\hat{H}_I&=&\sum\limits_{k=1}^N \hbar^2\,g_{k}\,
\left(\hat{n}_s+\frac{1}{2}\right)\,\left(\hat{n}_k+\frac{1}{2}\right).\label{HInt}
\end{eqnarray}
\end{subequations}
The constants $g_s$ and $g_k$ denote, respectively, the nonlinearity parameter and the coupling parameter, whereas $\omega_{s}$ and $\omega_k$ stand for the harmonic frequency of the nonlinear oscillator and the harmonic frequencies of the bath oscillators. The initial joint quantum state is assumed to be
\begin{eqnarray}\label{rhoQ0}
\hat{\rho}(0)=|\alpha_0\rangle\langle\alpha_0|\otimes\frac{e^{-\beta\hat{H}_R}}{\textrm{Tr}\left[e^{-\beta\hat{H}_R}\right]},
\end{eqnarray}
where $\beta=(k_B T)^{-1}$ and $T$ is the equilibrium temperature. The assumptions of separability (for the initial joint state) and purity (for the system state) are both convenient for a suitable construction of the respective classical distribution. The corresponding classical dynamics is described in terms of the Hamiltonian
\begin{subequations}
\begin{eqnarray}
H=H_S+H_R+H_I,
\end{eqnarray}
where
\begin{eqnarray}
H_S&=&\omega_s\,\left(\frac{p_s^2+q_s^2}{2} \right)+g_s\,\left(\frac{p_s^2+q_s^2}{2} \right)^2,\\
H_R&=&\sum\limits_{k=1}^{N}\omega_k\,\left(\frac{p_k^2+q_k^2}{2} \right),\\
H_I&=&\sum\limits_{k=1}^{N}g_k\left(\frac{p_s^2+q_s^2}{2}  \right)\left(\frac{p_k^2+q_k^2}{2}  \right).
\end{eqnarray}
\end{subequations}
For convenience, new canonical variables $q_k=X_k\sqrt{m_k\omega_k}$ and $p_k=P_k/\sqrt{m_k\omega_k}$ have been used instead of the usual position-momentum pair $(X_k,P_k)$. The quantum and classical Hamiltonians are correctly related by the scheme of bosonic quantization proposed in ref.\cite{angelo03}. A Gaussian spectral distribution given by
\begin{eqnarray}\label{gk}
g_k=\frac{\Omega}{\sqrt{N}}\,\exp\left[-\frac{\pi\,\left(
k-k_0\right)^2}{2\,N^2} \right],
\end{eqnarray}
where $k_0=N/2$ and $\Omega$ is constant, will be assumed in our calculations in the next sections. As has been shown in ref.\cite{angelo06} it allows for the appropriated implementation of the thermodynamic limit ($N\to\infty$). For simplicity and convenience, the assumption $\omega_k=\omega_0$ will be also adopted here. As discussed in ref.\cite{angelo06}, although restrictive, this choice does not change the main properties of the model, namely, the Gaussian decay for the quantum coherences and the structural form of the decoherence time. 

For the Liouvillian dynamics the following phase space distribution
is used:
\begin{eqnarray}\label{rho0}
\rho(0)=\frac{e^{-\frac{(q_s-q_0)^2}{\hbar}}}{\sqrt{\pi\,\hbar}}\frac{e^{-\frac{(p_s-p_0)^2}{\hbar}}}
{\sqrt{\pi\,\hbar}}\,\prod\limits_{k=1}^N\left(
\frac{e^{-\frac{p_k^2+q_k^2}{z\,\hbar}}}{z\,\pi\,\hbar}\right),
\end{eqnarray}
where
\begin{eqnarray}
z^{-1}=\tanh\left(\frac{\beta\,\hbar\,\omega}{2}\right).
\end{eqnarray}
This distribution corresponds to the Wigner function of $\hat{\rho}(0)$ given by \eqref{rhoQ0}. Note that the Wigner function for the reservoir state reduces to the traditional Boltzmanian function $e^{-\beta\,H_R}/\textrm{Tr}\left(e^{-\beta\,H_R} \right)$ only in the high-temperature regime $(\beta\,\hbar\,\omega\ll 1)$.

There are several reasons supporting the choice of this model, namely: (i) as it has been shown in \cite{angelo06}, the phase-damping reservoir induces nondissipative decoherence at finite temperature, this being one important difference from other purely diffusive environments \cite{adelcio,toscano}; (ii) the nonlinear oscillator used as the system of interest constitutes a paradigm of nontrivial quantum dynamics which exhibits interference and cat formation \cite{milburn,lili}; (iii) classical and quantum solutions are analytical and exact for all regimes of parameters, with no need for either Markovian or weak coupling assumptions.

\section{Analytical results}
\label{results}

\subsection{Newtonian results}

The Newtonian solution is obtained by integrating the system of $2(N+1)$ differential equations defined by Hamilton's equations, $\dot{q}_k=\partial_{p_k}H$ and $\dot{p}_k=-\partial_{q_k}H$. The final result for the Newtonian trajectory of the system can be conveniently written in matrix notation as follows:
\begin{subequations}\label{RNall}
\begin{eqnarray}\label{RN}
R_N(\tau)=\mathbb{M}_1\left[\phi_N\right]\,R_0,
\end{eqnarray}
where
\begin{eqnarray}
\phi_N(\tau)&\equiv&\frac{\omega_s\,\tau}{\hbar\,g_s}+\left(\frac{R_0^{\textrm{T}}
R_0}{\hbar}\right)\,\tau+\sum\limits_k \frac{g_k}{g_s}\,
\left(\frac{p_k^2+q_k^2}{\hbar}\right)\,\tau,\nonumber \\ \\
 \mathbb{M}_1[\chi]&\equiv&\left[
\begin{array}{cc}
\cos\chi & \sin\chi \\ \\ -\sin\chi &\,\,\,\cos\chi
\end{array}
\right],\label{7c}
\end{eqnarray}
\end{subequations}
where $R_N(\tau)\equiv\left( {q_s(\tau)\atop p_s(\tau)}\right)$, $R_0\equiv\left( {q_0\atop p_0}\right)$, and $R_0^{\textrm{T}}R_0=p_0^2+ q_0^2$. For convenience, all results have been written in terms of the following dimensionless parameter:
\begin{eqnarray}\label{tau}
\tau=\hbar\,g_s\,t.
\end{eqnarray}
Note that the Newtonian solution does not depend at all on $\hbar$. The effect of the nonlinearity becomes evident in equations above: the rotation matrix $\mathbb{M}[\phi_N]$ becomes dependent on the initial condition $R_0$. This fact can be interpreted as the basic mechanism responsible for the twist found in the dynamics of classical distributions \cite{milburn,adelcio} and for the dynamical formation of cat states (quantum superposition of several coherent states), as shown in ref.\cite{lili}.

\subsection{Quantum results}

The analytical solution for the reduced density matrix associated with
the system of interest is obtained by taking the trace of the global
density matrix, $\hat{\rho}(t)=e^{-\imath \hat{H}t/\hbar}\,\hat{\rho}(0)\,e^{\imath \hat{H}t/\hbar}$, over the reservoir degrees of freedom, i.e., $\hat{\rho}_S(t)=\textrm{Tr}_R\hat{\rho}(t)$. As shown in ref.\cite{angelo06}, the result can be written in Fock basis as
\begin{eqnarray}\label{rhotau}
&\hat{\rho}_S(\tau)&= \sum\limits_{n,n'=0}^{\infty}
e^{-|\alpha_0|^2}\frac{\alpha_0^n}{\sqrt{n!}}
\frac{(\alpha_0^{*})^{n'}}{\sqrt{n'!}}\, e^{-\imath (n^2-n'^2) \tau}\times\nonumber \\
&& e^{-\imath(n-n')\left(1+ \frac{\omega_s}{\hbar g_s} +\frac{G_1}{2
g_s} \right)\,\tau} \,C_{n-n'}(\tau)\,|n\rangle\langle n'|,
\end{eqnarray}
where $G_1\equiv\sum_k g_k$ and
\begin{eqnarray}
C_{n-n'}(\tau)\equiv\prod\limits_{k=1}^{N}\frac{1-e^{-\beta \hbar\,
\omega_k}}{1-e^{-\beta \hbar\,\omega_k} e^{-\imath
(g_k/g_s)\,(n-n')\, \tau}}.\label{C}
\end{eqnarray}
From \eqref{C} we see that
\begin{eqnarray}\label{|C|}
\left|C_{n-n'}
\right|=\prod\limits_k\left\{1+\frac{\sin^2\left[\frac{g_k}{g_s}\frac{(n-n')\,\tau}{2}
\right]}{\sinh^2\left[\frac{\beta \hbar\,\omega_k}{2}\right]}
\right\}^{-\frac{1}{2}}.
\end{eqnarray}
For short times $(\tau\ll 1)$ this function behaves as
\begin{eqnarray}\label{Gdecay}
\left|C_{n-n'} \right|\simeq e^{-(n-n')^2\,\tau^2/\tau_{DQ}^2},
\end{eqnarray}
where $\tau_{DQ}$, which denotes the characteristic decoherence time of the phase-damping reservoir, is given by
\begin{eqnarray}\label{tauDQ}
\tau_{DQ}=\sqrt{8}\,\frac{g_s}{G_2}\,\sinh\left(\frac{\beta\,\hbar\,\omega}{2}\right),
\end{eqnarray}
with $G_2^2\equiv\sum_k g_k^2$. (In the thermodynamic limit $G_2\simeq 0.89\,\Omega$.) Result \eqref{Gdecay} shows the character of Gaussian decay induced by the phase-damping reservoir over the off-diagonal terms ($n-n'\neq 0$) of the density operator given by \eqref{rhotau}. Hereafter the assumption $\omega_k=\omega$, as suggested in \cite{angelo06}, will be implemented for simplicity.

The column matrix $R_Q(\tau)\equiv\left({\langle \hat{q}_s
\rangle\atop \langle \hat{p}_s\rangle} \right)$, defined in terms of the
expectation values of the position and the momentum, is calculated
with the help of the usual relations: $\langle
\hat{q}_s\rangle=\sqrt{\hbar/2}(\langle \hat{a}^{\dag}_s\rangle+\langle
\hat{a}_s\rangle)$ and $\langle
\hat{p}_s\rangle=\imath\sqrt{\hbar/2}(\langle
\hat{a}^{\dag}_s\rangle-\langle \hat{a}_s\rangle)$, where $\langle
\hat{a}_s\rangle=\textrm{Tr} [\hat{a}_s\hat{\rho}_S(\tau)]$ and $\langle
\hat{a}^{\dag}_s\rangle=\textrm{Tr}
    [\hat{a}^{\dag}_s\hat{\rho}_S(\tau)]$. The calculations of these terms are too lengthy (though straightforward) in virtue of the many degrees of freedom of the system, and so they will be omitted. The final result can be written as
\begin{subequations} \label{RQall}
\begin{eqnarray}
R_{Q}(\tau)=\Big(\Gamma_{1Q}\,\mathbb{M}_1[\theta_{1Q}]\Big)\,\Big(\Lambda_{1Q}\,\mathbb{M}_1[\phi_{1Q}]\Big)\,R_0,\,\,\,\,\,\,\label{RQ}
\end{eqnarray}
where
\begin{eqnarray}
\Gamma_{1Q}(\tau)&\equiv&\left|C_{-1}(\tau)
\right|, \label{14b} \\ \nonumber \\
\Lambda_{1Q}(\tau)&\equiv&\exp\left[-\left(\frac{R_0^{\textrm{T}}
R_0}{\hbar}\right)\sin^2
\tau \right],\label{Lambda1Q}\\ \nonumber \\
\theta_{1Q}(\tau)&\equiv&\imath\,\ln\left[\frac{|C_{-1}(\tau)|}{C_{-1}(\tau)}
\right]+ \frac{G_1\,\tau}{2\,g_s}, \\ \nonumber \\
\phi_{1Q}(\tau)&\equiv& 2\,\tau
+\frac{\omega_s\,\tau}{\hbar\,g_s}+\left(\frac{R_0^{\textrm{T}}
R_0}{2\,\hbar}\right)\,\sin(2\,\tau).\label{14e}\qquad
\end{eqnarray}
\end{subequations}
The definition of the auxiliary dimensionless functions, $\Gamma_{1Q}$, $\Lambda_{1Q}$, $\theta_{1Q}$, and $\phi_{1Q}$, in equations \eqref{14b}-\eqref{14e}, allowed to write the final result \eqref{RQ} in such a suitable compact form, which will be useful for further comparisons. Interestingly, in Eq.\eqref{RQ} the term $\Lambda_{1Q} \mathbb{M}_1[\phi_{1Q}]$, depending solely on parameters associated with the quartic oscillator, and the term $\Gamma_{1Q} \mathbb{M}_1[\theta_{1Q}]$, related to the interaction with the reservoir, have been factorized. Note that $\Gamma_{1Q}$ is an attenuation factor describing the environmental decoherence (see Eq.\eqref{Gdecay}) whereas the factor $\Lambda_{1Q}$ is the one responsible for the well known revivals occurring in the closed dynamics of the quartic oscillator \cite{lili,milburn,agarwal}. In fact, in the absence of the reservoir ($\Omega=0$ and $\Gamma_{1Q} \mathbb{M}_1[\theta_{1Q}]=\mathbf{1}$), one may verify from \eqref{Lambda1Q} the occurrence of revivals at the instants
\begin{eqnarray}\label{tauR}
\tau_R=n\,\pi\qquad (n=1,2,3,\cdots).
\end{eqnarray}
This phenomenon, which has been shown to be a signature of the self-interference mechanism \cite{lili}, has no classical analogue, as can be seen in the corresponding Liouvillian result given by Eq.\eqref{Lambda1L}. In references \cite{adelcio,milburn}, numerical simulations comparing quantum (Wigner and Husimi) with classical distributions indeed demonstrate the absence of interference in the classical dynamics.

Note by \eqref{RNall} and \eqref{RQall} that one may rigorously verify that
\begin{eqnarray}\label{limQN}
\lim\limits_{\hbar\to 0} R_Q=R_N
\end{eqnarray}
in the absence of environmental reservoir ($\Omega=0$). In contrast, for $\Omega\neq 0$ this relation no longer holds, what can be demonstrated from \eqref{|C|} by showing that $\lim\limits_{\hbar\to 0}|C_{n-n'}|< 1$, for $t>0$. It is important to realize that this is not an effect of the many degrees of freedom composing the reservoir. In fact, the inequality applies even for only one oscillator in the reservoir. Rather, it occurs because the entanglement among the subsystems does not vanish asymptotically as $\hbar$ tends to zero (see, e.g., \cite{angeloK05}).

For the quadratic terms, $R_{2Q}\equiv\left({\langle \hat{q}^2_s\rangle \atop \langle \hat{p}^2_s\rangle}\right)$, a really tedious but equally straightforward calculation yields
\begin{subequations}
\begin{eqnarray}
&R_{2Q}(\tau)=\left(\frac{R_0^{\textrm{T}} R_0
+\hbar}{2}\right)\left({1\atop
1}\right)+&\nonumber \\
&\Gamma_{2Q}\,\Lambda_{2Q}\left({1\atop -1}
\right)\,\frac{R_0^{\textrm{T}}}{\sqrt{2}}\,\mathbb{M}_2\left[
\theta_{2Q}+\phi_{2Q}\right]\,\frac{R_0}{2},& \label{R2Q}
\end{eqnarray}
where
\begin{eqnarray}
\Gamma_{2Q}(\tau)&\equiv&\left|C_{-2}(\tau)
\right|, \\ \nonumber \\
\Lambda_{2Q}(\tau)&\equiv&\exp\left[-\left(\frac{R_0^{\textrm{T}}
R_0}{\hbar}\right)\sin^2 (2\,\tau) \right],\\ \nonumber \\
\theta_{2Q}(\tau)&\equiv&\imath\,\ln\left[\frac{|C_{-2}(\tau)|}{C_{-2}(\tau)}
\right]+\frac{G_1\,\tau}{g_s}, \\ \nonumber \\ \phi_{2Q}(\tau)&\equiv&
6\,\tau +\frac{2\,\omega_s\,\tau}{\hbar\,g_s}+\frac{R_0^{\textrm{T}}
R_0}{2\,\hbar}\,\sin(4\,\tau), \qquad
\\ \nonumber \\
\mathbb{M}_2[\chi]&\equiv&\left[
\begin{array}{cc}
\cos\chi & \sin\chi \\
\\ \sin\chi & -\cos\chi
\end{array} \right].
\end{eqnarray}
\end{subequations}
Once again, several dimensionless functions have been defined to guarantee compactness of the main result \eqref{R2Q}. Also, $\Gamma_{2Q}$ and $\Lambda_{2Q}$ are terms associated with the reservoir and with the quartic oscillator, respectively.

Then, for the quantum variances, written as
\begin{eqnarray}
\left(\Delta R_Q\right)^2\equiv\left({\Delta q_Q^2 \atop \Delta p_Q^2}
\right)=\left({\langle \hat{q}^2_s\rangle-\langle
\hat{q}_s\rangle^2\atop \langle \hat{p}^2_s\rangle-\langle
\hat{p}_s\rangle^2} \right),
\end{eqnarray}
we have
\begin{eqnarray}\label{DRQ}
&\left(\Delta R_Q\right)^2=\left({1\atop 1}
\right)\left[\frac{\hbar}{2}+\frac{R_0^{\textrm{T}}R_0}{2}\left(
1-\Gamma_{1Q}^2\,\Lambda_{1Q}^2\right) \right] +& \nonumber \\
&\left({1\atop -1}
\right)\frac{R_0^{\textrm{T}}}{\sqrt{2}}\Big\{\Gamma_{2Q}\Lambda_{2Q}\,\mathbb{M}_2[\psi_{2Q}]-
\Gamma_{1Q}^2\Lambda_{1Q}^2\,\mathbb{M}_2[2\psi_{1Q}] \Big\}
\frac{R_0}{\sqrt{2}},&
\nonumber \\
\end{eqnarray}
where $\psi_{iQ}\equiv\phi_{iQ}+\theta_{iQ}$, with $i=1,2$.

\subsection{Liouvillian results}

The time evolved Liouvillian distribution is obtained by integrating the Liouville equation, as shown in ref.\cite{angelo04}. The formal solution is given by
\begin{eqnarray}\label{21}
\rho(q_k,p_k,t)=\rho(q_k(-t),p_k(-t),0).
\end{eqnarray}
This formula establishes that the result can be obtained by replacing the arguments $q_k$ and $p_k$ of the initial distribution $\rho(q_k,p_k,0)$ given by Eq.\eqref{rho0} by the Newtonian trajectories evolved backwards in time, $q_k(-t)$ and $p_k(-t)$, respectively.  

Classical statistical averages are calculate as
\begin{eqnarray}\label{22}
\langle F \rangle_L(t)=\int dV\,F\,\rho(q_k,p_k,t),
\end{eqnarray}
where $dV=dq_sdp_sdq_1dp_1\cdots dq_Ndp_N$.

With formulas \eqref{21} and \eqref{22}, it is possible do calculate
the Liouvillian averages $\langle q_s\rangle$, $\langle q^2_s\rangle$,
$\langle p_s\rangle$, $\langle p^2_s\rangle$, and also the related
variances, $\Delta q_L^2=\langle q^2_s\rangle-\langle q_s\rangle^2$
and $\Delta p_L^2=\langle p^2_s\rangle-\langle p_s\rangle^2$. As for
the quantum analysis, although straightforward, the calculations are
too lengthy and do not add any further information to the results, so
they can be omitted. The Liouvillian results may be written as:
\begin{subequations}\label{RLall}
\begin{eqnarray}\label{RL}
 R_L(\tau)=\Big(\Gamma_{1L}\,\mathbb{M}_1[\theta_{1L}]\Big)\,\Big(\Lambda_{1L}\,
\mathbb{M}_1[\phi_{1L}]\Big)\,\mathbb{N}\,R_0,\,\,\,\,\,
\end{eqnarray}
where
\begin{eqnarray}
D_m(\tau)&\equiv&\prod\limits_{k}\left[1-\imath\,m\,\left(\frac{g_k}{g_s}\right)\,\frac{z\,\tau}{2}\right]^{-1},\label{22b}
\\ \nonumber \\
\Gamma_{1L}(\tau)&\equiv&\left|D_1(\tau) \right|,\label{22c} \\ \nonumber \\
\eta_1(\tau)&\equiv&1+\tau^2,\\ \nonumber \\
\Lambda_{1L}(\tau)&\equiv&\eta_1^{-1}\exp\left\{-\left(\frac{R_0^{\textrm{T}}R_0}{\hbar}\right)\frac{\tau^2}{\eta_1(\tau)}
\right\},\label{Lambda1L}\qquad\\ \nonumber \\
\theta_{1L}(\tau)&\equiv&\imath\,\ln\left[\frac{|D_1(\tau)|}{D_1(\tau)}
\right],
\\ \nonumber \\ \phi_{1L}(\tau)&\equiv&\frac{\omega_s\,\tau}{\hbar\,
g_s}+\frac{\tau}{\eta_1(\tau)}\,\left(\frac{R_0^{\textrm{T}}R_0}{\hbar}\right),\\
\nonumber \\ \mathbb{N}(\tau)&\equiv&\frac{1}{\eta_1}\left[
\begin{array}{cc}
1- \tau^2 & 2\,\tau \\
\\-2\,\tau & 1- \tau^2
\end{array} \right],\,\,\,\,\,\,
\end{eqnarray}
\end{subequations}
with
$\mathbb{N}^{\textrm{T}}\mathbb{N}=\mathbb{N}\mathbb{N}^{\textrm{T}}=\mathbf{1}$.
$\mathbb{M}_1$ is the rotation matrix defined by \eqref{7c},
$R_L\equiv\left({\langle q_s \rangle \atop \langle p_s\rangle}
\right)$, and $R_0=\left({q_0\atop p_0}\right)$. In Eq.\eqref{22b},
the index $m$ is such that $m=1,2$ (see, e.g., equations \eqref{22c} and \eqref{25b}).

In the short-time regime the classical attenuation factor
reduces to
\begin{eqnarray}
\left|D_m(\tau)\right|&=&e^{-m^2\,\tau^2/\tau_{DL}^2},
\end{eqnarray}
where $\tau_{DL}$, the Liouvillian analogue of the decoherence time \eqref{tauDQ}, is given by
\begin{eqnarray}\label{tauDL}
\tau_{DL}=\sqrt{8}\,\frac{g_s}{G_2}\,\tanh\left(\frac{\beta\,\hbar\,\omega}{2}\right).
\end{eqnarray}

Once again the reservoir contributions could be factorized in the analytical solution for the centroid. Here, however, an extra rotation associated with the nonlinearity of the oscillator is imposed upon the initial condition $R_0$ through the matrix $\mathbb{N}$.

It is worth emphasizing that the attenuation factor $\Lambda_{1L}$
does not exhibit any revival, as has been mentioned. While
quantum distribution are able to suffer self-interference, which in
turn produces relocalization and revival in the amplitude of the
centroid, classical distributions do not interfere, as shown in \cite{adelcio,milburn}.

For the quadratic terms, $R_{2L}\equiv\left({\langle
q_s^2\rangle\atop \langle p_s^2\rangle} \right)$, we have
\begin{subequations}
\begin{eqnarray}
&R_{2L}(\tau)=\left(\frac{R_0^{\textrm{T}} R_0
+\hbar}{2}\right)\left({1\atop
1}\right)+&\nonumber \\
&\Gamma_{2L}\,\Lambda_{2L}\left({1\atop -1}
\right)\,\frac{R_0^{\textrm{T}}}{\sqrt{2}}\,\mathbb{M}_2\left[
\theta_{2L}+\phi_{2L}\right]\,\mathbb{N}_2\,\frac{R_0}{\sqrt{2}},
\,\,\,\,\,\,\,\,\,\,\,\,\,\,\,\, \label{R2L}
\end{eqnarray}
where
\begin{eqnarray}
\Gamma_{2L}(\tau)&\equiv&\left|D_{2}(\tau)
\right|, \label{25b}\\ \nonumber \\
\Lambda_{2L}(\tau)&\equiv&\eta_2^{-1}\exp\left[-4\left(\frac{R_0^{\textrm{T}}
R_0}{\hbar}\right)\frac{\tau^2}{\eta_2(\tau)} \right],\\ \nonumber \\
\eta_2(\tau)&\equiv&1+4\,\tau^2, \\ \nonumber \\
\theta_{2L}(\tau)&\equiv&\imath\,\ln\left[\frac{|D_2|}{D_2} \right], \\ \nonumber \\
\phi_{2L}(\tau)&\equiv&
\frac{2\,\omega_s\,\tau}{\hbar\,g_s}+2\left(\frac{R_0^{\textrm{T}}
R_0}{\hbar}\right)\, \frac{\tau}{\eta_2}, \qquad\\
\nonumber \\ \mathbb{N}_2&\equiv&\frac{1}{\eta_2^2}\left[
\begin{array}{cc}
1-12\,\tau^2 & 2\,\tau\,(3-4\,\tau^2) \\
\\ -2\,\tau\,(3-4\,\tau^2) & 1-12\,\tau^2
\end{array} \right].\qquad
\end{eqnarray}
\end{subequations}

For the Liouvillian variances, written as
\begin{eqnarray}
\left(\Delta R_L\right)^2\equiv\left({\Delta q_L^2 \atop \Delta p_L^2}
\right)=\left({\langle q^2_s\rangle-\langle q_s\rangle^2\atop
\langle p^2_s\rangle-\langle p_s\rangle^2} \right),
\end{eqnarray}
the following result has been obtained:
\begin{eqnarray}\label{DRL}
&\left(\Delta R_L\right)^2=\left({1\atop 1}
\right)\left[\frac{\hbar}{2}+\frac{R_0^{\textrm{T}}R_0}{2}\left(
1-\Gamma_{1L}^2\,\Lambda_{1L}^2\right) \right]+
\left({1\atop -1}\right) \times &\nonumber \\ &\frac{R_0^{\textrm{T}}}{\sqrt{2}}\Big\{\Gamma_{2L}\Lambda_{2L}\,\mathbb{M}_2[\psi_{2L}]\,\mathbb{N}_2-  
 \Gamma_{1L}^2\Lambda_{1L}^2\,\mathbb{M}_2[2\psi_{1L}]\,\mathbb{N}_1^2
\Big\} \frac{R_0}{\sqrt{2}},&\nonumber \\
\end{eqnarray}
where $\psi_{iL}\equiv\phi_{iL}+\theta_{iL}$ with $i=1,2$.

\begin{figure}[ht]
\centerline{\includegraphics[scale=0.46]{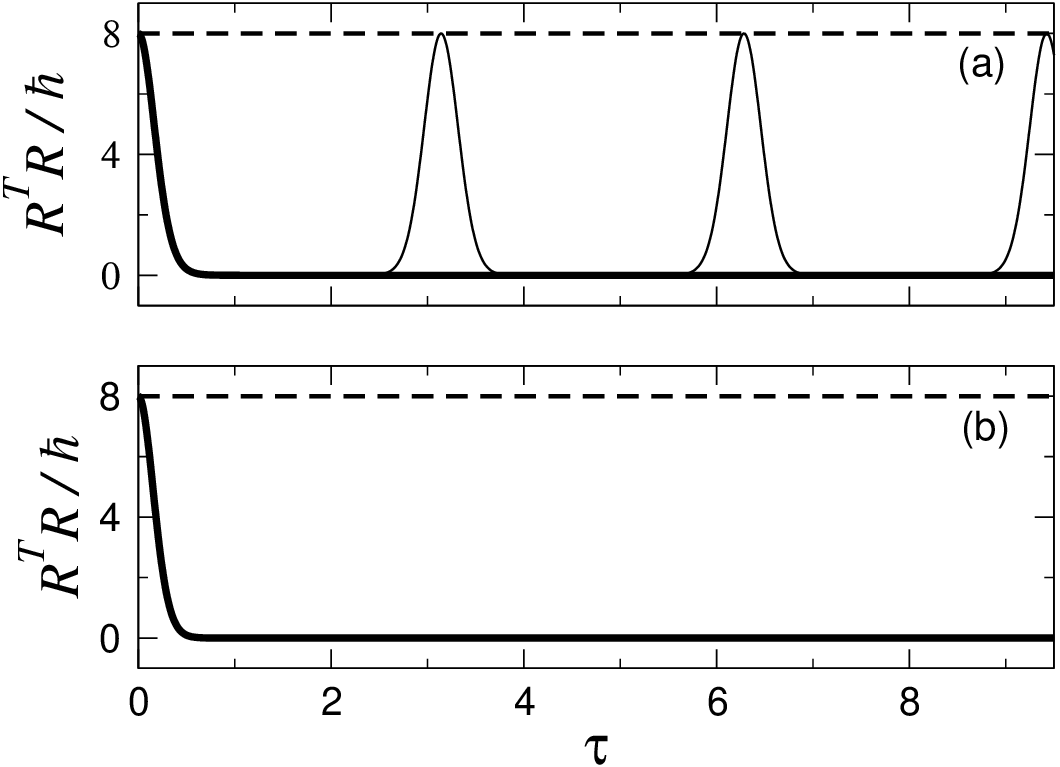}}
\caption{Behavior of $R^{\textrm{T}}(\tau) R(\tau)/\hbar$ as a function of $\tau$ for Newtonian (dashed lines), quantum (thin solid lines) and Liouvillian (thick solid lines) theories. These numerical results have been obtained for $R_0^{\textrm{T}}R_0/\hbar=8.0$ and $g_s/\Omega=5.0$ (weak coupling). In (a) it is shown the simulations for the closed dynamics ($\Omega=0$), for which the revival phenomenon can be observed in the quantum dynamics at the instants $n\pi$. In (b) we see the results for the open dynamics ($\Omega=0.1$ and $N=50$) in a regime of high temperature ($\beta\hbar\,\omega=0.1$). Quantum result becomes classical, in the Liouvillian sense, due to environmental decoherence.}
\label{fig1}
\end{figure}

\begin{figure}[ht]
\centerline{\includegraphics[scale=0.46]{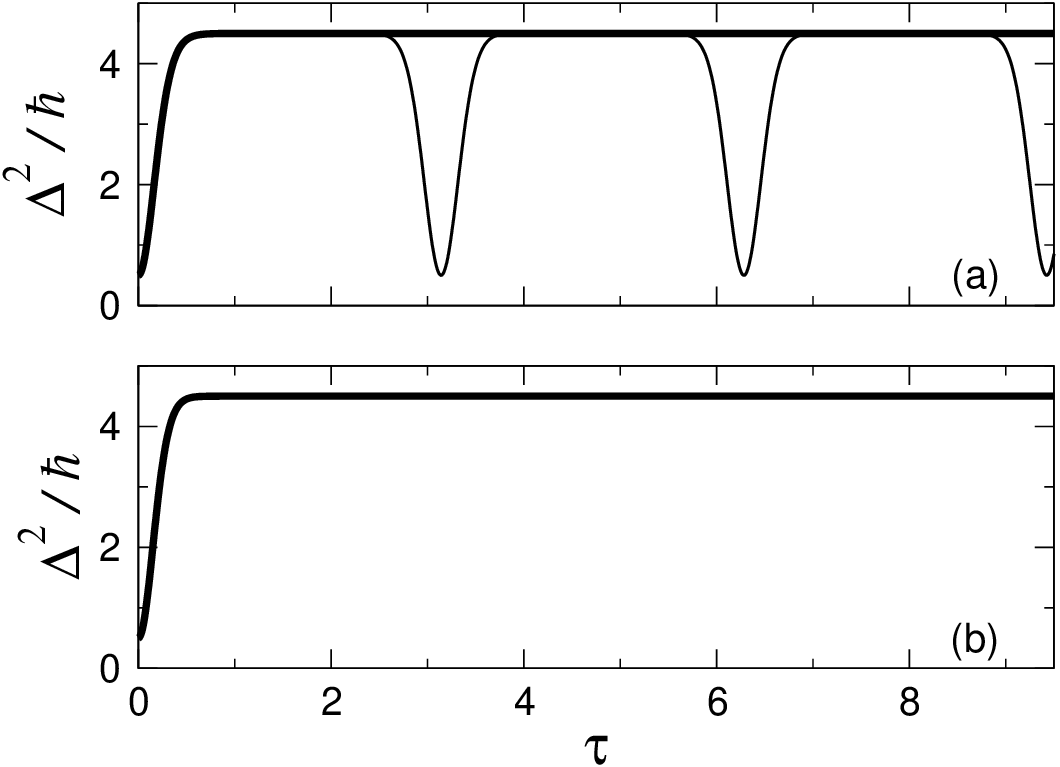}}
\caption{Behavior of the mean variance defined by $\Delta^2\equiv(\Delta q^2+\Delta p^2)/2$ as a function of $\tau$ for quantum (thin solid line) and Liouvillian (thick solid line) theories in units of $\hbar$. These numerical results have been obtained for the same parameter values of Fig.\ref{fig1}. The simulations for the closed dynamics ($\Omega=0$) are shown in (a), where it is possible to see the relocalization of the quantum wave function at the instants $n\pi$. In (b) we see the results for the open dynamics ($\Omega=0.1$ and $N=50$) in a regime of high temperature $(\beta\hbar\,\omega=0.1)$. Decoherence induces diffusion which makes the quantum variances similar to the Liouvillian ones.}
\label{fig2}
\end{figure}

Figures \ref{fig1} and \ref{fig2} illustrates the main characteristics of quantities such as $R_N^{\textrm{T}}R_N=R_0^{\textrm{T}}R_0$ (Newtonian), $R_Q^{\textrm{T}}R_Q=\Gamma_{1Q}^2\Lambda_{1Q}^2R_0^{\textrm{T}}R_0$ (quantum), and $R_L^{\textrm{T}}R_L=\Gamma_{1L}^2\Lambda_{1L}^2R_0^{\textrm{T}}R_0$ (Liouvillian) for the closed ($\Omega=0$) and the open ($\Omega> 0$) dynamics of the quartic oscillator. Quadratic terms such as $R^{\textrm{T}}R$, constructed with the results \eqref{RN}, \eqref{RQ}, and \eqref{RL}, are convenient to plot because they do not display the oscillations due to the rotation matrix $\mathbb{M}_1$. The effect of diffusive decoherence in destructing quantum revivals becomes evident in the figure. Next, the correspondence principle is discussed with the help of these results.

\section{Quasi-Determinism}
\label{QNC}

The concept of {\em determinism} is mainly concerned with philosophical beliefs, according to which there are laws (hidden or not) univocally connecting the past and the future of every dynamical system, including animals and human beings. There is no chance for free choice. Note that the success of Newtonian theory in making predictions for the macroscopic world gives scientific support for such a philosophical concept. On the other hand, any attempt to apply the concept to the microscopic world will fail, once the notion of a trajectory linking the past and the future does not exist in the quantum realm. In fact, variances are always present, as predicted by Heisenberg's uncertainty principle. Despite this fact, it is still possible to defend the idea of determinism by speculating on the existence of some sort of hidden laws (or {\em hidden variables}). 

This paper, however, does not intend to settle this controversial question from an ontological viewpoint, asserting at the end whether determinism exist or not. Rather, the aim is to use the tools of the physical science, namely, fundamental laws and experimental investigations, to evaluate under which conditions the concept of determinism is acceptable in physics. In particular, the following test will be employed here: if the position and the momentum of a particle can be correctly predicted at a given instant {\em for an individual system}, then the concept of determinism will be regarded as a meaningful one in physics. This is what we call a scenario of {\em quasi-determinism}, since the motion will be experimentally indistinguishable from the truly deterministic Newtonian one. It is important to keep in mind this specific connotation attributed here to the word {\em determinism}, since it will be used from now on to incite the discussion about the central problem: the correspondence principle.

The argument motivating the historical statement of the correspondence principle is the one according to which quantum results must match classical ones because macroscopic world is correctly described by classical physics. However, from a rigorous point of view this reasoning cannot be considered as completely correct. Actually, it strongly depends on our capability of making the experimental checking. Extremely accurate measurements on the center of mass position of a macroscopic projectile near the surface of the Earth, would indicate some dispersion around the Newtonian parabolic trajectory. In fact, once we admit that the classical world is governed by quantum laws at a microscopic level, we have to conceive the center of mass motion as being determined by the quantum dynamics of all quantum particles composing the projectile. Accordingly, the center of mass is expected to behave quantum mechanically, with variances such as $\Delta x_{cm}=(\langle \hat{x}_{cm}^2\rangle-\langle \hat{x}_{cm}\rangle^2)^{\frac{1}{2}}$ being associated with its motion. That is, in principle, the quantum nature of the center of mass, and thus the deviation from Newtonian predictions, could be verified by such an accurate measurements. In this context, the status of the Newtonian mechanics as the basis for enunciating the correspondence principle is questionable.

To circumvent these difficulties, the following approach is proposed. We assume that the quantum-classical correspondence is achieved when quantum predictions are consistent with the concept of determinism (such as defined above). That is, if quantum mechanics is capable of making correct predictions for an individual system (what is expected to occur in a macroscopic regime) then the correspondence principle will be satisfied. As a consequence, since in this case the wave function dispersion is expected not to be detectable in virtue of the insufficient experimental resolution, Newtonian predictions can be successfully applied as well. Then, the approach is the following: for insufficient experimental accuracy, quantum dispersion is not apparent, Newtonian predictions is adequate, and the concept of deterministic behavior survives in physics. In this case, in virtue of Ehrenfest's theorem, quantum mechanics will also succeed in making predictions for individual systems, at least for the short-time dynamics of initially localized wave functions (see, e.g., the famous example of the chaotic moon of Saturn \cite{zurek98,zurek96}, for which the estimated Ehrenfest time is about two decades) and, at the end, Einstein's requirements will be satisfied.

Let us consider, as a first example, the quantum dynamics of a free particle initially prepared in a minimal uncertainty state. A well known calculation \cite{cohen} shows that the product of the quantum variances evolves as
\begin{eqnarray}\label{28}
\Delta x(t)\,\Delta
p(t)=\frac{\hbar}{2}\sqrt{1+\left(\frac{\hbar\,t}{2\,m\,\Delta
x^2_0}\right)^2}.
\end{eqnarray}
Consider now an experience in which the phase space resolution, defined by the measurement device, is given by the action $\delta S$, which here will be given by $\delta S=M\,\hbar/2$, $M$ being an arbitrary real number. The product given in \eqref{28} achieves the critical value $\delta S$ for
\begin{eqnarray}
t_{det}=t_E\,M\sqrt{1-\frac{1}{M^2}},
\end{eqnarray}
where
\begin{eqnarray}
t_{E}=\frac{2\,m\,\Delta x_0^2}{\hbar}.
\end{eqnarray}
$t_{E}$ is the time scale firstly derived by Ehrenfest
\cite{ehrenfest} for the critical spreading of the free particle
wave function. (In his work, Ehrenfest showed that $t_E$ can reach large values for macroscopic particles.) Now, within the time scale defined by $t_{det}$ the quantum dispersion of the wave function is smaller than the experimental resolution, so it is not possible to distinguish the quantum character of the phenomenon. In this case, Newtonian mechanics can be successfully applied to make predictions about the future phase space state $(x,p)$ of the particle. In other words, up to the instant  $t_{det}$ the idea of a deterministic behavior is physically acceptable (within the experimental precision considered). Then, within this time scale, quantum mechanics satisfies Einstein's principle as a theory capable of making predictions for individual systems, and in turn, the quantum-classical correspondence (as we have defined it) has been achieved. Beyond this time scale, quantum mechanics indeed fails in making predictions for individual systems, but exactly the same thing occurs with Newtonian mechanics. The determinism has been lost and the correspondence breaks off.

Now these ideas are applied to the dynamics of the nonlinear oscillator in the
presence of the reservoir. Is decoherence capable of extending the
time scale of determinism? It has already been pointed out in some
works \cite{wiebe,ballentine} that decoherence is not able to reduce
the wave function spreading. Since the spreading is the basic
mechanism responsible for the breakdown in the determinism, as it
has been suggested by the example above, one can anticipate that
decoherence is not able to realize such a task. In fact
that can be demonstrated in our model as follows.

Firstly, it is easy to see that decoherence does not restrain
spreading. Using $g_s=0$ (the system of interest is now a harmonic
oscillator) one obtains, by \eqref{Gdecay} and \eqref{DRQ}, the
following long-time ($\tau>\tau_{DQ}$) variances product:
\begin{eqnarray}
\Delta q_Q(\infty)\,\Delta
p_Q(\infty)=\frac{\hbar}{2}+\frac{R_0^{\textrm{T}}R_0}{2}.
\end{eqnarray}
This result is larger than $\hbar/2$, which is the value obtained for the closed dynamics of a harmonic oscillator. Clearly, purely diffusive decoherence has induced diffusion.

Secondly, let us consider the nonlinear oscillator under decoherence
in a situation in which the experimental resolution is written as $\delta S=M\hbar/2$. Since $q$ and $p$ plays essentially the same
role in our model we chose for convenience an average phase space
variance $\Delta^2\equiv (\Delta q_Q^2+\Delta p_Q^2)/2$
instead of the product $\Delta q_Q \Delta p_Q$. Defining the determinism break time, $\tau_{det}$, as the instant at which the mean variance becomes comparable to the experimental resolution ($\Delta^2=\delta S$), one may use \eqref{DRQ} to obtain
\begin{eqnarray}\label{32}
1-\Gamma_{1Q}^2(\tau_{det})\,\Lambda_{1Q}^2(\tau_{det})=\frac{\hbar}{R_0^{\textrm{T}}R_0}(M-1).
\end{eqnarray}
Interestingly, a quite similar result is obtained when one compares the quantum centroid \eqref{RQ} with the Newtonian trajectory \eqref{RN} through the prescription $\frac{1}{2}(R_N^{\textrm{T}}R_N-R_Q^{\textrm{T}} R_Q)=\delta S$. This attests the consistence of our approach: the determinism break time, $\tau_{det}$, which corresponds to the time of critical spreading, is directly related to the instant at which quantum centroid and Newtonian trajectory stop agreeing.

Expanding Eq.\eqref{32} for $\tau_{det}\ll 1$ one obtains
\begin{eqnarray}\label{taudet}
\tau_{det}=\frac{\tau_{det}^{(c)}}{\sqrt{1+\left[\frac{\tau_{det}^{(c)}}{\tau_{DQ}}\right]^2\frac{2
R_0^{\textrm{T}}R_0}{(M-1)\hbar}}},
\end{eqnarray}
where
\begin{eqnarray}
\tau_{det}^{(c)}=\sqrt{\frac{(M-1)\hbar^2}{2(R_0^{\textrm{T}}R_0)^2}}.
\end{eqnarray}
In these expressions $\tau_{det}^{(c)}$ denotes the break time for the {\em closed} dynamics of the nonlinear oscillator and $\tau_{DQ}$ the decoherence time given by Eq.\eqref{tauDQ}. Equation \eqref{taudet} demonstrates that decoherence does not extend the determinism time scale, since $\tau_{det}<\tau_{det}^{(c)}$. Therefore, in the context of Einstein's requirements for individual systems, decoherence plays no essential role.

In addition, by \eqref{tau} one may write $\tau_{det}=\hbar\,g_s\,t_{det}$ to
show, with the help of the results above, that $t_{det}\propto
\hbar^{-\frac{1}{2}}\propto t_{E}$, where $t_E$ is the Ehrenfest time for the quartic oscillator \cite{angelo03,berman81}. This is another indicative that the time scale of determinism is indeed equivalent to the Ehrenfest one.

\section{The quantum to classical transition}
\label{QLC}

Once we accept the validity of the uncertainty principle at the basis of the microscopic phenomena we are forced to regard the Newtonian mechanics as an incomplete theory. In fact, since the initial phase space state $(q_0,p_0)$ of a particle can never be determined with arbitrary precision, the future phase space state can be predicted only statistically. Hence, the feeling of an objective reality disappears. In this scenario, the Liouvillian theory turns out to be the more general classical theory, whereas the Newtonian theory works as an approximation for the short-time regime of narrow distributions, as suggested in the precedent section. Thus, the correspondence principle is now discussed by comparing quantum with Liouvillian results, for times beyond the determinism time scale defined in the precedent section.

Firstly one should note from our analytical and exact results that there is no regime of parameters (with or without the reservoir) capable of making quantum and classical results mathematically identical. Although apparently naive, this sentence emphasizes, in agreement with \cite{adelcio,wiebe,ballentine}, the approximated character of the quantum-classical correspondence and, in turn, the role of the experimental resolution in defining it. Accordingly, once again the theoretical action $\delta S=M\hbar/2$, with $M$ real, will play the role of the experimental phase space resolution in our calculations. 

The future phase space state of an individual system subjected to either quantum or classical statistical fluctuations can be predicted only in terms of averages and variances as $(\langle q\rangle,\langle p\rangle)\pm \frac{1}{2}(\Delta q,\Delta p)$. Then, the quantum-classical correspondence will be investigated in terms of the following differences:
\begin{eqnarray}\label{DR}
D(\tau)=\frac{1}{2}\left|\langle
R\rangle_Q^{\textrm{T}}(\tau)\,\langle R\rangle_Q(\tau)-\langle
R\rangle_L^{\textrm{T}}(\tau)\,\langle R\rangle_L(\tau)\right|
\end{eqnarray}
for the centroids and
\begin{eqnarray}\label{DV}
D(\tau)=\left|\frac{\Delta q_Q^2(\tau)+\Delta
p_Q^2(\tau)}{2}-\frac{\Delta q_L^2(\tau)+\Delta
p_L^2(\tau)}{2}\right|
\end{eqnarray}
for the variances.

These definitions disregard, for simplicity, the influence of
the rotation matrices $\mathbb{M}_{1,2}$ and $\mathbb{N}$ but capture the
attenuation effects due to both the reservoir and the nonlinearity. 
Using results \eqref{RQ}, \eqref{RL},
\eqref{DRQ}, and \eqref{DRL}, its is possible to show that both
equations above reduce to
\begin{eqnarray}
D(\tau)=\left|\Gamma_{1Q}^2(\tau)\,\Lambda_{1Q}^2(\tau)-\Gamma_{1L}^2(\tau)\,\Lambda_{1L}^2(\tau)\right|\,
\frac{R_0^{\textrm{T}}R_0}{2}.
\end{eqnarray}
This result demonstrating the equivalence between \eqref{DR} and \eqref{DV} suggests the existence in this model of a common dynamics for the quantum-classical differences associated with the centroids and with the variances. As a consequence, one may conceive the existence of a single time scale describing the breakdown in the correspondence, this being a counter-example contributing to the discussion raised in ref.\cite{adelcio03}.

\begin{figure}[t]
\centerline{\includegraphics[scale=0.45]{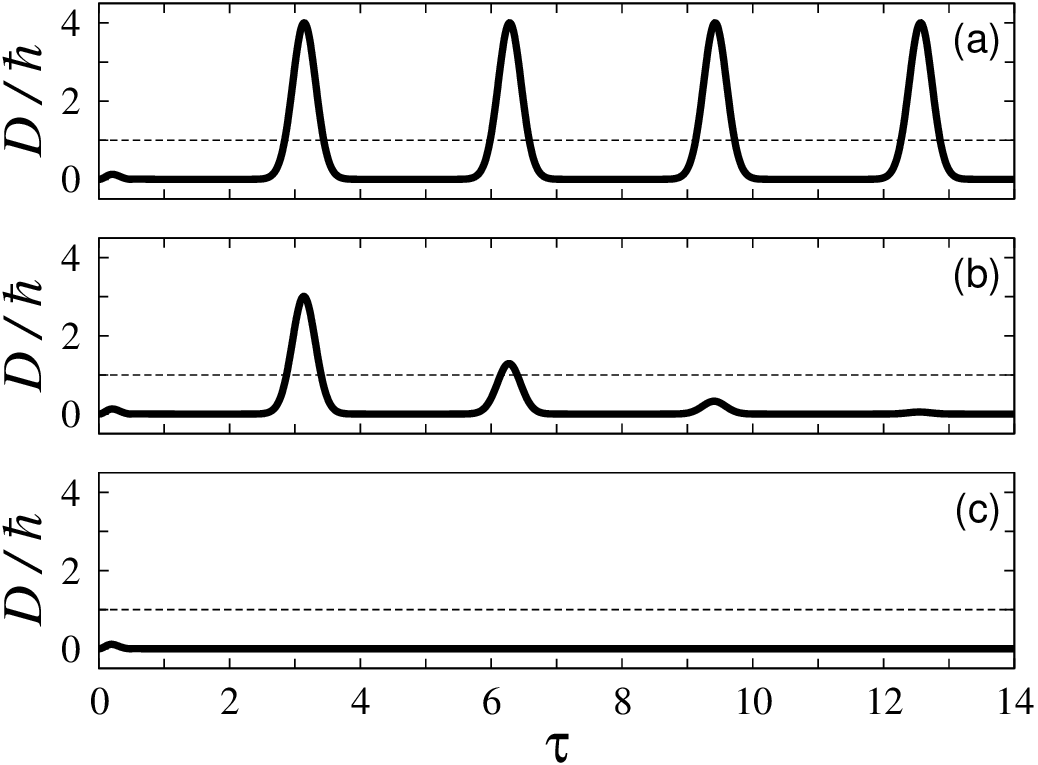}}
\caption{Numerical results of $D(\tau)/\hbar$ obtained for the same parameter values of Fig.\ref{fig1}. The horizontal dashed lines indicate the critical resolution in units of $\hbar$ (here, $\delta S/\hbar=1$). In (a), where the simulation for the closed dynamics ($\Omega=0$) is plotted, the differences occurs due to the quantum revival. In (b) and in (c) we see the results for the open dynamics ($\Omega=0.1$ and $N=50$), for which $\beta\,\hbar\,\omega=1.0$ in (b) and $\beta\,\hbar\,\omega=0.1$ in (c). Note that quantum results become classical only in the high-temperature regime.}
\label{fig3}
\end{figure}

In Fig.\ref{fig3} the behavior of the function $D$ is shown as a function of $\tau$ for the closed and the open dynamics of the nonlinear oscillator in two different regimes of temperature. Note that diffusive decoherence, modelled in terms of our phase-damping reservoir, is indeed effective in restoring the quantum-classical correspondence. However, a word concerning the role of temperature is in order.

Results \eqref{tauDQ} and \eqref{tauDL} indicate that the attenuation effects due to the reservoir may occur also for low temperatures. However, in such a regime the differences between quantum and Liouvillian decay times become more accentuated. Indeed, from \eqref{tauDQ} and \eqref{tauDL} it is easy to show that
\begin{eqnarray}
\tau_{DQ}=\tau_{DL}\,\cosh\left(\frac{\beta\,\hbar\,\omega}{2} \right).
\end{eqnarray}
Now, the differences in the decays for the low-temperature regime ($\beta\,\hbar\,\omega\gg 1$) become evident. This result turns out to be an indicative that the high-temperature regime is an additional {\em necessary} condition for the establishment of the quantum-classical transition. However, this is not the whole truth, since the condition $\tau_{DQ}=\tau_{DL}$ does not automatically imply that $\tau_{DQ}< \tau_R$, where $\tau_R$ is the revival time given by \eqref{tauR}. In fact, since $\tau_R$ is the time scale which determines the appearance of purely quantum effects, such as quantum interference, the classical limit is guaranteed by imposing that decoherence occurs early. In general one may state that the quantum-classical transition is achieved as long as the decoherence time be smaller than the time scale for which quantum phenomenon (e.g., self-interference) takes place in the dynamics. In our model this condition implies, by \eqref{tauDQ} and \eqref{tauR}, that
\begin{eqnarray}
\frac{k_B\,T}{\hbar\,\omega}> \frac{g_s}{2\,\Omega},
\end{eqnarray}
where the high-temperature regime ($\beta\,\hbar\,\omega\ll 1$) and the thermodynamic limit ($N\to\infty$, $G_2\to0.89\,\Omega$) have already been considered. This simple relation establishes the lower bound for the value of the reservoir temperature capable of yielding the quantum-classical transition. In addition, this result shows that nondissipative decoherence inducing quantum-classical transition can occur at finite temperatures.

\section{Summary}

In this paper, the exact quantum and Liouvillian dynamics of a quartic oscillator coupled with a purely diffusive reservoir at arbitrary temperature have been solved analytically in terms of expectation values and variances associated with the phase space variables. The results demonstrate the effectiveness of our reservoir model in describing nondissipative decoherence and inducing quantum-classical transition.

Concerning the correspondence principle, this paper intend to contribute in two main directions. On one hand, examples were given showing that the Ehrenfest time is intimately connected with the time scale within which the concept of determinism (suitably defined) is acceptable in physics. In fact, within the Ehrenfest time scale initially localized wave functions remain sufficiently narrow in such a way that its dispersion cannot overcome the limits imposed by the experimental resolution. In this case, Newtonian and quantum mechanics are both well succeeded in making predictions for individual systems, thus satisfying Einstein's requirements. Beyond the Ehrenfest time scale quantum predictions deviate from Newtonian ones, but the latter are no longer correct. This analysis gives answer to Einstein's insatisfaction and also attributes a new physical meaning to the Ehrenfest time. In addition, it has been shown that diffusive decoherence plays no essential role within such a time scale, this being a point of agreement with the claims of ref.\cite{wiebe,ballentine}.

On the other hand, in a second direction, this paper shows that although decoherence is indeed necessary to reestablish quantum-classical correspondence beyond the determinism time scale, a certain amount of coarse-graining (related to experimental resolution) is also required in order to characterize the correspondence in a more precise way. This is necessary since the quantum-classical correspondence cannot be fully achieved at a mathematical level. Also, our results points out to the fact that the high-temperature regime is another fundamental ingredient in the scenario of the quantum-classical correspondence. A quantification of how high the reservoir temperature must be, in order to guarantee the correspondence, is given for our model in terms of a lower bound for the temperature. Although the assumption of $T\to \infty$ appears in several approaches (see, e.g., \cite{adelcio,toscano}) as an artifact to simulates the effects of diffusive decoherence from the traditional dissipative master equations, here it is justified in terms of physical conditions. Yet, our approach points out to the possibility of modelling diffusive decoherence at finite temperatures.

\acknowledgments

The author would like to thank A. F. Gomes, A. D. Ribeiro, C. F. Woellner, F. R. L. Parisio and A. S. Sant'Anna for fruitful discussions and helpful suggestions.


\end{document}